\documentstyle[psfig]{mn2e}

\def\etal{{\it et al. }}
\title[Globular Clusters in NGC 1052]
{Evolutionary History of the Elliptical Galaxy NGC 1052}

\author[Pierce, Brodie, Forbes, Beasley, Proctor \& Strader] {
  Michael Pierce$^{1}$\thanks{mpierce@astro.swin.edu.au}, 
  Jean P. Brodie$^{2}$,
  Duncan A. Forbes$^{1}$,
   Michael A. Beasley$^{1,2}$,\\
\\
\LARGE
  Robert Proctor$^{1}$, Jay Strader$^{2}$\\
  $^1$ Centre for Astrophysics \& Supercomputing, Swinburne University, Hawthorn, VIC 3122, Australia\\
  $^2$ Lick Observatory, University of California, Santa Cruz, CA 95064, USA\\
}
\begin{document}
\maketitle  

\begin{abstract}

We have obtained Keck spectra for 16 globular clusters (GCs)
associated with the merger remnant elliptical NGC~1052, as well as a
long-slit spectrum of the galaxy.  We derive ages, metallicities and
abundance ratios from simple stellar population models using the
methods of Proctor \& Sansom (2002), applied to extragalactic GCs for
the first time.  A number of GCs indicate the presence of strong blue
horizontal branches which are not fully accounted for in the current
stellar population models. We find all of the GCs to be $\sim 13$ Gyr
old according to simple stellar populations, with a large range of
metallicities.  From
the galaxy spectrum we find NGC~1052 to have a luminosity-weighted
central age of $\sim 2$ Gyr and metallicity of [Fe/H]$\sim +0.6$. No
strong gradients in either age or metallicity were found to the
maximum radius measured (0.3 r$_{e} \simeq 1$ kpc).  However, we do
find a strong radial gradient in $\alpha$--element abundance, which
reaches a very high central value. The young central starburst age is
consistent with the age inferred from the HI tidal tails and infalling
gas of $\sim 1$ Gyr.  Thus, although NGC~1052 shows substantial
evidence for a recent merger and an associated starburst, it appears
that the merger did not induce the formation of new GCs, perhaps
suggesting that little recent star formation occurred. This
interpretation is consistent with ``frosting'' models for early-type
galaxy formation.
 
\end{abstract}


\section{Introduction}\label{sec_intro}

Globular clusters (GCs) are generally considered to be good tracers of
a galaxy's star formation history.  Bimodal colour distributions are
observed in many GC systems, suggesting that galaxies undergo multiple
epochs of star and GC formation (Harris 2001).  The exact nature of
how these multiple star formation events occur is crucial to
understanding galaxy formation.

Ashman \& Zepf (1992) describe the production of metal-rich GCs during
the merger of two gas-rich disc galaxies.  This scenario results
in similar ages for the metal-rich GCs and the merger event.  In this
model the blue, metal-poor GCs which belonged to the progenitor
galaxies are universally old ($\sim 13$ Gyr).  Forbes, Brodie \&
Grillmair (1997) propose a multi-phase collapse, during which
metal-poor GCs form early during the pre-galaxy collapse phase and, at
a later time, metal-rich GCs form out of more enriched gas at a
similar time to the galaxy field stars.  This model implies a
metal-rich population that is a few Gyr younger than the metal-poor
population.  The exact age difference may depend on mass and
environment (Beasley \etal 2002).  On the other hand, Cote, Marzke \&
West (1998) suggest the bimodality of GC colours in large ellipticals
is due to the gradual accretion of metal-poor GCs from dwarfs, with
metal-rich GCs being indigenous to the elliptical.  In this picture,
the metallicity of the metal-rich population is a function of the
galaxy's luminosity and both sub-populations should be old.

These physical processes may all occur to some extent.  To distinguish
between the relative contributions of each process, we need examples
of elliptical galaxies covering a range of masses and in different
environments.  Given that the above GC formation scenarios have different
expectations concerning the ages of the metal-poor and metal-rich
sub-populations, determining the age of individual GCs is a crucial
step in testing these models.

There are several methods to measure GC properties such as age and
metallicity from integrated spectra.  Examples include the method of
Brodie \& Huchra (1990; hereafter BH90), who present an empirical
metallicity measure based on absorption line features.  Another method
is that of Strader \& Brodie (2004; hereafter SB04), based on a
principal components analysis (PCA) of 11 absorption line features.
Methods involving spectral indices, predominantly the Lick system, and
their comparison to simple stellar population (SSP) models can break
the age-metallicity degeneracy (Worthey 1994).  These methods usually
rely on an age sensitive index (e.g., H$\beta$) and a metallicity
sensitive index (e.g., Mgb) to break the degeneracy.  However, these
methods suffer difficulties when individual indices are contaminated,
due to cosmic rays, skyline subtraction or galactic emission. Another
complication is non-solar $\alpha$-abundance ratios which are
difficult to determine from index-index plots and can affect the
higher order Balmer indices (Thomas, Maraston \& Korn 2004).
Recently, Proctor \& Sansom (2002; hereafter PS02) have used a
multi-index $\chi^{2}$--minimisation method to obtain ages,
metallicities and abundance ratios for the stellar populations of late
and early-type galaxies.  Another advantage of using many indices is
that it maximises the information used to break the age-metallicity
degeneracy.  Here we extend this method to extragalactic GCs for the
first time (see Proctor, Forbes \& Beasley 2004 for an application to
Galactic GCs).

Hot luminous blue horizontal branch (BHB) stars have a significant
effect on the measurement of several indices (most notably the Balmer
indices, see Lee, Yoon \& Lee 2000). Studies of Galactic GCs show that
those with similar ages and metallicities can have very different
spectral contributions from their HB stars depending on whether the HB
stars are hot (blue) or cool (red). The way the HB is modelled has a
significant impact on a SSP model (Maraston, Greggio \& Thomas 2001a).
A colour-magnitude diagram is probably the best way to deduce HB
morphology in a given GC, but such diagrams are limited to GCs in the
very nearest galaxies.  Thus information on HBs in extragalactic GCs
must generally come from integrated spectra.  Old metal-rich GCs with
BHBs (e.g., NGC 6388 and NGC 6441; Rich \etal 1997) can falsely appear
to be intermediate-age clusters due to this strengthening of the
Balmer lines (Schiavon \etal 2004), therefore, the ability to deduce
HB morphology from integrated spectra is important. Schiavon \etal
(2004) have recently introduced a method to detect anomalous HBs in GC
spectra by differential comparisons of H$\delta$, H$\gamma$, and
H$\beta$ line strengths.

There is evidence for a significant number of proto-GCs in galaxies
which are currently merging (e.g., The Antennae; Whitmore \& Schweizer
1995) and very recent ($<$500 Myr) mergers such as NGC 7252 (Maraston
\etal 2001b; Schweizer \& Seitzer 1998; Miller \etal 1997) and NGC
3921 (Schweizer, Seitzer \& Brodie 2004; Schweizer \etal 1996). The
outstanding questions are whether a substantial fraction of the
proto-GC sub-population formed in these mergers survives, and hence
whether it can account for the GC systems of ellipticals (see Forbes,
Brodie \& Grillmair 1997).

An important link between ongoing mergers and ``old'' elliptical
galaxies is the intermediate aged (2-5 Gyr) merger remnants. Previous
GC spectroscopic studies of merger remnant galaxies such as NGC 1316
(Goudfrooij \etal 2001) and NGC 3610 (Strader \etal 2003; Strader,
Brodie \& Forbes 2004) have confirmed a small number of GCs with ages
matching the time since the merger event ($\sim 3$ Gyr and $\sim 2$
Gyr respectively).  However, the total number of intermediate-age GCs
in these galaxies is currently poorly constrained spectroscopically.

In this work we examine the GC system of NGC~1052, which is an
excellent candidate for an elliptical that has undergone a minor merger
event in the last few Gyr.  NGC~1052 is an E4 galaxy, located in a
small group at a distance of 18 Mpc (Forbes, Georgakakis \& Brodie
2001b).  Aside from containing an active nucleus, NGC~1052 displays
several indications of a past merger or significant accretion event.
The gas angular momentum is higher than that of the stellar component,
with the gas and stars having different rotation axes (van Gorkom
\etal 1986, Plana \& Boulesteix 1996).  NGC~1052 also reveals HI
'tidal tails' (van Gorkom \etal 1986), infalling HI gas onto the
active nucleus (van Gorkom \etal 1989) and dust lanes (Forbes, Sparks
\& Macchetto 1990) all signs of an accretion or merger.  van Gorkom
\etal (1986) suggest that the HI observations could be explained by
the accretion of a gas-rich dwarf, or other minor interaction, about 1
Gyr ago.

Despite the evidence for a recent accretion event, NGC~1052 shows
almost no optical disturbance (it has a low fine structure value of
$\Sigma$ = 1.78, Schweizer \& Seitzer 1992) and has a fundamental
plane residual of +0.07 (Prugniel \& Simien 1996) which is consistent
with a normal galaxy on the fundamental plane (Forbes, Ponman \& Brown
1998).

We can further probe the evolutionary history of NGC~1052 by measuring
the ages and metallicities of individual GCs in the galaxy.  The
imaging of Forbes \etal (2001b) revealed a bimodal GC system as is
typical for elliptical galaxies. The two sub-populations had a colour
difference of $\Delta$B--I=0.4$\pm$0.1. Using Worthey (1994) models,
under the assumption that the blue GC sub-population is 15 Gyr old and
metal-poor with [Fe/H]=--1.5, Forbes \etal (2001b) showed that any
newly-formed red GCs (assuming ages of $\le$2.5 Gyr) must have supersolar
metallicities to fully account for the red sub-population.

In Section \ref{sec_obs} we will present our observations and data
reduction methods.  Measured indices and our analysis of these Lick
indices are given in Section \ref{sec_inds}.  Using the method of
PS02, metallicities, ages and abundance ratios are given in Section
\ref{sec_aam}.  Spectra of NGC~1052 itself are given in Section
\ref{sec_long} and a brief analysis of GC kinematic data is in Section
\ref{sec_gal}.  Finally we discuss the implications of our results in
Section \ref{sec_disc} and present conclusions in Section
\ref{sec_conc}.

\section{Observations and data reduction}\label{sec_obs}

Spectra of GC candidates around NGC~1052 were obtained with the Low
Resolution Imaging Spectrometer (LRIS; Oke \etal 1995) on the Keck I
telescope. Candidate selection, based on the Keck imaging data of
Forbes \etal (2001b), was designed to cover a wide range of potential
GC colours (i.e. 1.2 $<$ B--I $<$ 2.5, 0.6 $<$ V--I $<$ 1.7).  The
properties of our candidates can be found in Table 1.  Observations
were obtained in 2003 January 25-27 with an integration time of $16 \times
1800$s = 8 hours for the slit-mask.  However, the signal-to-noise 
ratios of the
combined spectra were improved by excluding the exposures made at the
highest airmass. These suffered from signal-to-noise degradation,
especially at bluer wavelengths.  Seeing ranged between
0.65$^{''}$ and 0.75$^{''}$ over the three nights.  A 600 lines per mm
grating blazed at 4000 \AA\ was used for the blue side, resulting in
an approximate wavelength range of 3300 - 5900 \AA\ and a FWHM
spectral resolution of $\sim3.3$ \AA.

\begin{figure} 
\centerline{\psfig{figure=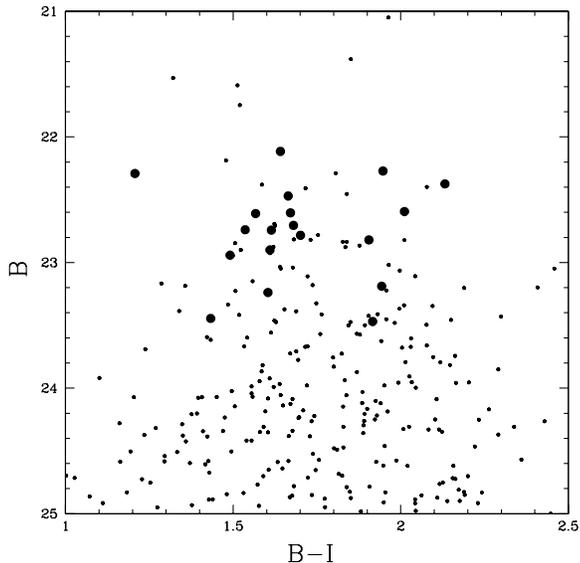,width=0.45\textwidth,angle=0}} 
\caption{Candidate globular cluster colour magnitude distribution.  GCs for which spectra were obtained are represented by large symbols.  The small symbols are the full sample from the imaging study of Forbes \etal (2001b). Spectra were obtained for luminous GC candidates from both the blue and red sub-populations.}
\label{n1052colour}     
\end{figure}

\begin{figure} 
\centerline{\psfig{figure=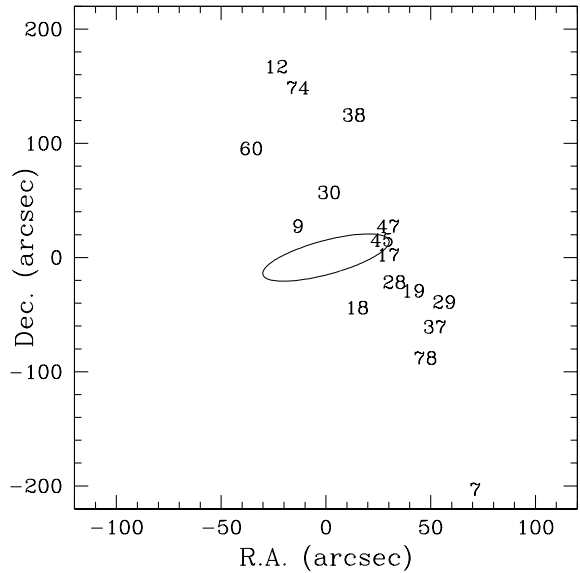,width=0.45\textwidth,angle=0}} 
\caption{Spatial distribution of the confirmed GCs.The numbers correspond to the 
GCs listed in Table 1. 
The E4 ellipse represents the effective radius of NGC~1052 at r$_e\sim$34''. The orientation is North up and East left. Note the x and y scales are different.}
\label{positions}     
\end{figure}

Table 1 presents observational data for the objects for which we
obtained Keck spectra.  Figure \ref{n1052colour}, a B--I colour
magnitude diagram, shows that we have obtained spectra from both the
blue and red sub-populations of GC candidates.  We have also generally sampled the most luminous GCs in NGC~1052.  If any young ($\sim$ few Gyrs old) GCs exist then they will tend to be brighter than the old population.  Plotted in Figure
\ref{positions} are the spatial positions of the confirmed GCs.

\begin{table*}
\begin{center}
\caption{\scriptsize Globular Cluster candidates around NGC~1052.  GC ID, R.A., Dec., V magnitude, B--I and V--I colours are from Forbes \etal (2001b). Galactocentric radii are calculated from GC positions assuming the distance to NGC~1052 of 18 Mpc. Heliocentric velocities are from this work.}
\renewcommand{\arraystretch}{1.5}
\begin{tabular}{lcccccrr}
\hline
\hline
ID & R.A.   & Dec.    & V    & B--I & V--I & Radius & V$_{helio}$\\
GC &(J2000) & (J2000) & (mag) & (mag) & (mag) & (kpc) & (km/s)\\
\hline
GC7  & 02:41:00.0 & -8:18:44.0 & 21.32 &  1.64   &  0.84   & 19.2 & 1659$\pm$74   \\
GC9  & 02:41:05.7 & -8:14:53.5 & 21.41 &  1.95   &  1.08   & 2.4  & 1649$\pm$105   \\
GC11 & 02:41:00.2 & -8:17:32.7 & 21.68 &  1.21   &  0.60   & ....  & star        \\
GC12 & 02:41:06.4 & -8:12:33.9 & 21.40 &  2.13   &  1.16   & 14.2  & 1610$\pm$123   \\
GC17 & 02:41:02.8 & -8:15:18.6 & 21.72 &  1.66   &  0.92   & 1.9  & 1498$\pm$101   \\
GC18 & 02:41:03.8 & -8:16:04.9 & 21.71 &  2.01   &  1.13   & 4.3  & 1583$\pm$96    \\
GC19 & 02:41:02.0 & -8:15:50.1 & 21.85 &  1.67   &  0.92   & 4.7  & 1640$\pm$92    \\
GC22 & 02:41:07.5 & -8:13:13.1 & 21.87 &  1.68   &  0.85   & ....  & star    \\
GC28 & 02:41:02.6 & -8:15:42.6 & 22.03 &  1.54   &  0.83   & 3.8  & 1414$\pm$86    \\
GC29 & 02:41:01.0 & -8:16:00.0 & 22.02 &  1.70   &  0.94   & 6.3  & 1273$\pm$78    \\
GC30 & 02:41:04.7 & -8:14:24.0 & 21.95 &  1.91   &  1.04   & 4.7  & 1537$\pm$122   \\
GC37 & 02:41:01.3 & -8:16:21.7 & 22.05 &  1.61   &  0.92   & 6.4  & 1443$\pm$76    \\
GC38 & 02:41:03.9 & -8:13:16.2 & 22.15 &  1.61   &  0.86   & 10.6  & 1295$\pm$122   \\
GC45 & 02:41:03.0 & -8:15:05.5 & 22.24 &  1.49   &  0.79   & 2.8  & 1765$\pm$74    \\
GC47 & 02:41:02.8 & -8:14:53.5 & 21.90 &  1.57   &  0.86   & 3.5  & 1781$\pm$83    \\
GC55 & 02:41:00.8 & -8:16:32.8 & 22.36 &  1.94   &  1.12   & ....  & galaxy\\
GC60 & 02:41:07.2 & -8:13:45.5 & 22.45 &  1.60   &  0.82   & 8.5  & 1415$\pm$122   \\
GC74 & 02:41:05.7 & -8:12:52.3 & 22.74 &  1.43   &  0.73   & 12.6  & 1675$\pm$93    \\
GC78 & 02:41:01.6 & -8:16:49.0 & 22.59 &  1.92   &  1.03   & 8.9  & 1619$\pm$92    \\
\hline
\end{tabular}
\label{tableprop}
\end{center}
\end{table*}

Data reduction was carried out using standard IRAF\footnote{IRAF is
distributed by the National Optical Astronomy Observatories, which are
operated by the Association of Universities for Research in Astronomy,
Inc., under cooperative agreement with the National Science
Foundation} methods.  Tracing of spectra was done using the two
exposures from each night with the lowest airmass (both $<1.2$).
These were combined to increase the signal for aperture tracing.  They
were then used as the reference apertures for their respective nights
and spectra were extracted from individual exposures.

Comparison lamp spectra of Hg, Ar, Ne, Cd and Zn were used for
wavelength calibration (mostly based on 8 Hg lines). Zero-point
corrections of up to 3 \AA\ were performed on the science spectra
using the bright OI skyline at 5577.34 \AA. Different methods of
spectral combining were tested.  Average combining was used with
median scaled pixel rejection, due to the significant
difference in flux levels between high and low airmass observations.
Sigma clipping was used to reject cosmic rays and reduce the effect of
strong skylines.  The resulting spectra have S/N = 18--45 \AA$^{-1}$
measured at 5000 \AA.

Flux calibrations were provided by the flux standards Feige 34 and
G191B2B. These were taken on the same run by long-slit and therefore
have slightly different wavelength coverage to the multi-slit spectra.
The velocities of GC candidates were determined by cross-correlation
against high signal-to-noise M31 GC templates, i.e. 225--280 and
158--213 (see Beasley \etal 2004).  The typical variation of
$V_{helio}$ with respect to the different templates is $\pm$ 20 km/s.

\begin{figure} 
\centerline{\psfig{figure=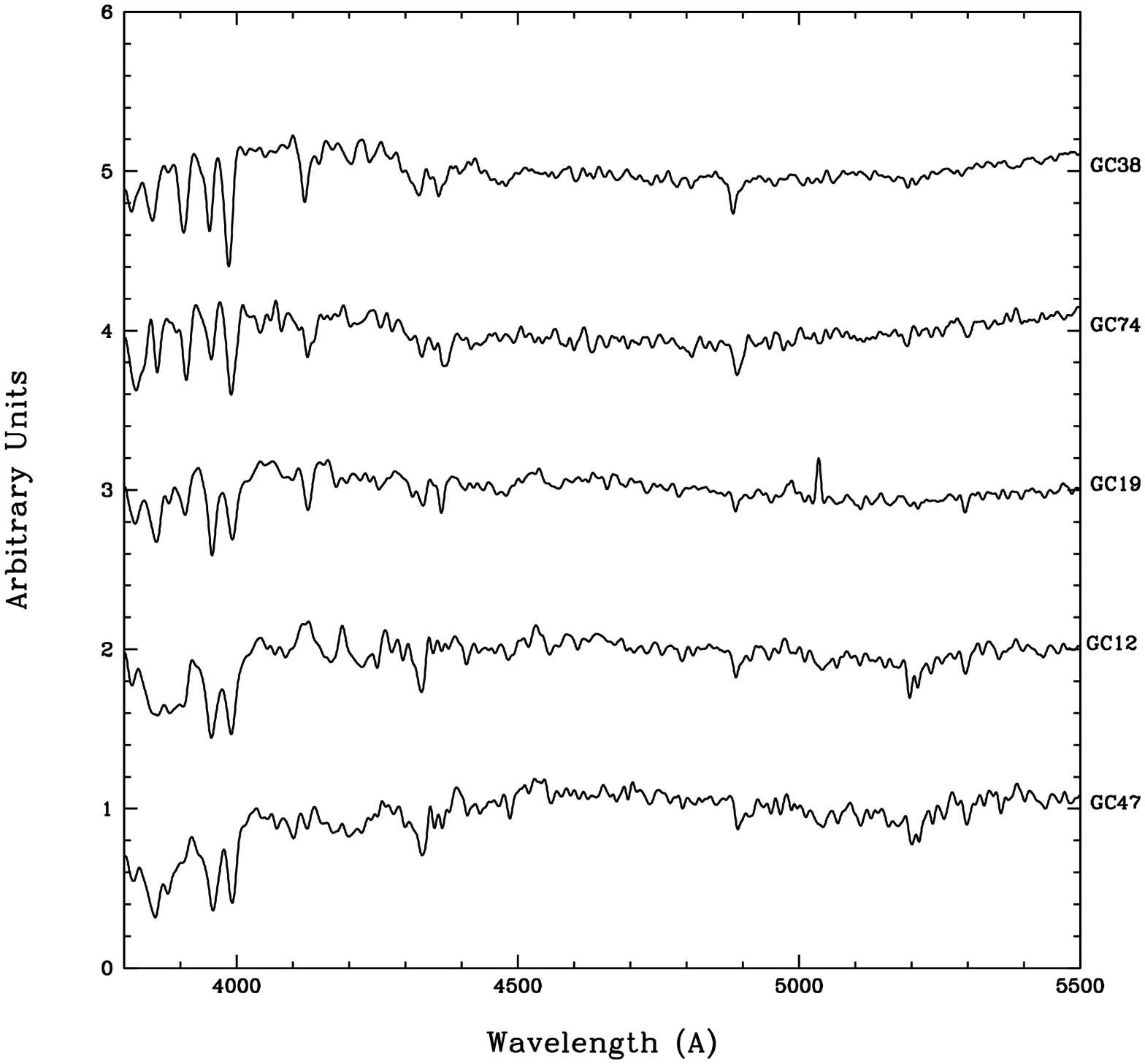,width=0.45\textwidth,angle=0}} 
\caption{Selected Keck spectra of NGC~1052 GCs normalized and offset
by 1 unit.  These spectra have not been de-redshifted (which
introduces a shift of $\sim$ 25 \AA). The spectra show the strong CaII H and
K features around 3950 \AA, the Gband at 4300 \AA, H$\beta$ (4863
\AA), Mg feature (5200 \AA), Fe lines around 5300 \AA\ and other
absorption lines.  The emission line at $\sim$5035 \AA\ in GC19 is
[OIII]5007 \AA\ from the galaxy which did not fully subtract. The spectra indicate a
range of metallicities and CaII ratios.}  
\label{spec}     
\end{figure}

Of the 19 spectra obtained, one is a background galaxy with z=0.204,
two are Galactic stars and the remaining 16 have velocities consistent
with being GCs associated with NGC~1052.  Thus, we have a success rate
of 16 out of 19.  Figure \ref{spec} shows the confirmed GC spectra
smoothed to the Lick resolution.  Each spectrum has been normalized,
then offset by a flux interval of one, to make visual comparison easier.  These
spectra show a wide variation in the ratio of the CaII H and K
features around 3950 \AA.  Other prominent lines are the G band at 4300
\AA, H$\beta$ (4863 \AA), Mg features (5200 \AA) and Fe lines around
5300 \AA.

\section{GC Spectral Line Indices}\label{sec_inds}

\subsection{Measurement of Indices}

We measured Lick indices (Trager \etal 1998; Worthey \& Ottaviani
1997), the Rose (1984) CaII index, and BH90 indices from our
flux-calibrated spectra after convolving the spectra with a
wavelength-dependent Gaussian kernel to broaden to the Lick
resolution.  We then applied small offsets to the Lick indices based
on measurements of several Lick standard stars (see Beasley \etal
2004).  Uncertainties in the Lick indices were derived from the photon
noise in the unfluxed spectra.  Errors on the H$\beta$ index range
from $\pm$ 0.21--0.36 \AA.  The Lick indices and Rose CaII index are
presented for the confirmed GCs in Tables 2 and 3.  We note that the
5202 \AA\ skyline falls in the central band for Mg$_{2}$ and Mgb for
our redshifted spectra, but in the Mgb side band for the Lick standard
stars, raising some doubt over the accuracy of the offsets applied to
the Mgb index and possible systematic errors in Mg$_{2}$.

Missing values in Tables 2 and 3 indicate problematic indices due to
various factors.  Ionised gas emission from NGC~1052 has contaminated
the H$\beta$ index in GCs 9 and 19. This can be seen clearly by
examining the [OIII] emission feature at 5007 \AA\ for GC19 in Figure
\ref{spec}.  Difficulty subtracting the 5202 \AA\ skyline from the Mg
indices (especially Mgb) affects GCs 45, 60 and 74.  The presence of
some weak light from internal reflections within the LRIS instrument
from the guide stars affects wavelengths below 4200~\AA\ of GCs 12,
30, 45 and 60.  Indices in parentheses are clipped from our fitting
process due to their large deviance from the best fit age, metallicity
and abundance ratio (see Section 3.3).

\begin{table*}
\begin{center}
\caption{\scriptsize Globular Cluster indices $\lambda <$ 4500 \AA.  Central index values (first line) and errors (second line).  Indices in brackets are removed during the fitting process.}
\renewcommand{\arraystretch}{1.5}
\begin{tabular}{lccccccccccc}
\hline
\hline
 ID & CaII & H$\delta_{A}$ & H$\delta_{F}$ & CN$_{1}$ &  CN$_{2}$  &  Ca4227  &  G band & H$\gamma_{A}$ &  H$\gamma_{F}$ & Fe4383  & Ca4455 \\
GC  & ratio & (\AA) & (\AA) & (mag) & (mag) & (\AA) & (\AA) & (\AA) & (\AA) &  (\AA) & (\AA) \\
\hline

GC7   & 1.01  &  2.07 &  2.60 &  (-0.039) &  -0.038 &  (0.08) &  (2.70) &  -0.06 &  0.83 & 1.08 & 0.73 \\  
      & 0.07 & 0.33 & 0.21  & 0.010 &  0.012 &  0.19 &  0.34 &  0.35 &  0.22 &  0.52 & 0.27 \\
GC9   & 1.41  &  0.06 & 0.70 & (0.061) &  (0.100) &  1.19 &   (3.28) &  -2.81 &  -0.15 &   3.72 &   1.00 \\
 & 0.17 & 0.33 &  0.23 &  0.009 &  0.011 &  0.16  & 0.29 & 0.34 & 0.21 & 0.44 & 0.23 \\
GC12  & 1.05  &  .... &  ....  &   ....  &  .... &  (0.86) &  (4.07) &  (-3.90) & (-1.04) & 3.20 & 1.44 \\
 & 0.10       & ....  & ....  & .... &  .... &  0.17  & 0.30  & 0.35  & 0.22  & 0.47 & 0.24 \\
GC17  & 0.94  &   2.81 &   2.20 &  (-0.057) & -0.050 &  (-0.33) &  2.55 &  0.44 &  1.16 &  -1.16 &  1.24 \\
 &	0.08 & 0.30  & 0.21  & 0.009 &  0.011 &  0.19  & 0.32  & 0.34 &  0.21  & 0.54 & 0.27 \\
GC18  & 1.05  &  -2.05 & -0.07 &  (0.010) &  0.017 & 1.09 & 4.08 & (-2.16) & (-0.03) &  3.37 &  1.34 \\
 &	0.13 & 0.41 &   0.27  & 0.011 &  0.013 &  0.21  & 0.36  & 0.42 &  0.26  & 0.57 & 0.28 \\
GC19  & 1.17  &  2.79 & (3.30) & (-0.024) &  (0.020) &  0.46 &  2.35 & -0.36 &  1.30 &  1.40 &  0.21 \\
 &	0.10 & 0.35  & 0.22  & 0.010  & 0.012  & 0.21  & 0.36  & 0.39 &  0.24  & 0.58 & 0.31 \\
GC28  & 1.20  &  0.99 &  1.81 &  (0.029) &  (0.049) &  0.63 &  3.98 & -3.01 & -0.11 &  2.31 &  1.00 \\
 &	0.15 & 0.39 &  0.25  & 0.011  & 0.013  & 0.21   & 0.38  & 0.42  & 0.26  & 0.62 & 0.31 \\
GC29  & 0.96  & 1.83 &  (3.17) & -0.061 & -0.022 & (-0.22) & (1.26) & (-0.26) &  1.21 &  2.66 &  1.12 \\
 &	0.10 & 0.37 &   0.23  & 0.011  & 0.013 &  0.23 &   0.38 &   0.40 &   0.24 &   0.58 & 0.31 \\
GC30  & ....  & ....  &   ....  &  ....  & ....  & 0.71 & 6.02 & -4.99 & -0.35 & 4.43 & 1.09 \\
 &	.... & .... &  .... & ....  &  ....  & 0.21 &   0.34 &   0.44 &   0.26 &   0.55 & 0.30 \\
GC37  & 0.94  &  2.23 & 1.89 & (-0.071) & -0.056 & 0.63 & 1.99 & -0.01 & 0.97 & 1.13 & 1.07 \\
 &	0.11 & 0.39 &  0.27  & 0.011 &  0.014  & 0.21 &   0.41 &   0.41 &   0.25 &   0.63 & 0.32 \\
GC38  & 0.78  & 3.68 & 3.01 & (-0.089) & (-0.048) & -0.26 & (1.45) & 1.42 & (1.40) & -1.13 & 0.50 \\
 & 0.07 & 0.34  & 0.23  & 0.010  & 0.012  & 0.22  & 0.38  &  0.38 &  0.24  & 0.62 & 0.32 \\
GC45  & 1.14  &  .... & ....   &   ....  &  .... & (-0.03) & (2.58) & 3.02 & 2.72 & -0.33 & 0.64 \\
 & 0.12 & .... &  ....  & ....  & .... &  0.29 & 0.48 &  0.47 &  0.29  & 0.68 & 0.35 \\
GC47  & 1.13  & -2.53 &  -0.11 &  (0.102) &  0.104 & (-0.39) &  (3.59) & (-2.23) & (0.69) & 4.48 & 1.48 \\
 & 0.13 & 0.40  & 0.26  & 0.010 &  0.013 &  0.20  & 0.36 &  0.39  & 0.23  & 0.52 & 0.28 \\
GC60  & 0.84  &  .... &   ....  &  ....  &  .... &  (-1.02) & (3.79) & 0.49 & 1.81 & 0.27 & (-1.58) \\
 & 0.10 & .... & ....  &  .... &   ....&  0.27  & 0.41 &  0.44  & 0.27 &  0.71 & 0.38 \\
GC74  & 0.81  &   3.98 & (2.40) & -0.107 & -0.083 &  0.36 &  0.27 &  2.55 &  1.96 &  1.94 &  0.61 \\
 & 0.09 & 0.41 &  0.29 &  0.013  & 0.015 &  0.25 &   0.48 &   0.46 &   0.28 &   0.73 & 0.39 \\
GC78  & 1.09  &  -0.03 &  1.21 & -0.009 &  0.011 &  0.67 & (-0.72) &  (0.84) &  (1.46) &  2.31 &  0.13 \\
 & 0.15 & 0.50 &  0.33 &  0.014  & 0.017 &  0.27 &   0.53 &   0.51 &   0.31 &   0.77 & 0.42 \\

\hline
\end{tabular}
\label{indsblue}
\end{center}
\end{table*}

\begin{table*}
\begin{center}
\caption{\scriptsize Globular Cluster indices $\lambda >$4500 \AA.  Central index values (first line) and errors (second line).  Indices in brackets are removed during the fitting process.}
\renewcommand{\arraystretch}{1.5}
\begin{tabular}{lcccccccccc}
\hline
\hline

ID & Fe4531  &  C4668   &  H$\beta$ &  Fe5015  &  Mg$_{1}$   &  Mg$_{2}$   &  Mgb    &  Fe5270  &  Fe5335  &  Fe5406\\
GC  & (\AA) & (\AA) & (\AA) & (\AA) & (mag) & (mag) & (\AA) & (\AA) & (\AA) & (\AA) \\
\hline

GC7   &   (2.84) &  1.34 &  2.50 &  3.10  &  (0.012) &  (0.074) &  (0.98) &  1.46 &  1.46 &  0.94 \\  
      &   0.42  &   0.65 &   0.24 &   0.53    &   0.005 &  0.006 &  0.25  & 0.27 &  0.31 & 0.23 \\
GC9   &   2.74 &  1.19 &  .... &  3.33  &  0.062 &  (0.182) &  2.80 &  2.14 &  (0.76) &  1.26 \\ 
      & 0.35    &   0.54 &  ....  &   0.45  &   0.004 &  0.005 &  0.21 &   0.23 &   0.26 & 0.19 \\
GC12  &  3.30 &  3.15 &  1.76 &  4.93 &  0.085 &  (0.235) &  3.56 &  2.02 &  1.97 &  1.05 \\
      & 0.36 &   0.57 &  0.22 &   0.47 &   0.005 &  0.006 &  0.22 &  0.24 &   0.27 &  0.20 \\
GC17  & 2.61 & (-2.46) &  2.34 &  (0.31) &  (0.043) &  0.089 &  0.91 &  0.42 & (-0.55) &  0.62 \\  
      & 0.41 &  0.67  & 0.25 &   0.56 &   0.006 &  0.006 &  0.27 &  0.30 &  0.34 &  0.24 \\
GC18  & 2.61 &  2.62 &  (2.34) &  3.78 &  0.087 &  (0.257) &  3.32 &  (2.87) &  1.27 &  0.64 \\  
      & 0.44 &   0.66 &   0.25 &   0.55 &   0.006 &  0.006 &  0.25 &  0.28 &  0.31 &  0.23 \\
GC19  & 2.23 &  0.72 &  .... &  1.53 &  (0.058) &  0.072 &  1.10 &  1.31 &  0.18 &  0.47 \\  
      & 0.46    & 0.72      &  ....   &  0.62 &  0.006  & 0.007  & 0.29 &  0.32 &  0.37 & 0.26 \\
GC28  & 2.25 &  (2.68) &  2.17 &  (2.42) &  0.062 &  (0.125) &  1.93 &  2.55 &  (0.48) &  1.22 \\  
      & 0.48  & 0.74 &  0.28 &  0.64  & 0.006 &  0.007 &  0.30 &  0.33 &  0.38 & 0.27 \\
GC29  &  1.97 &  (2.35) &  (2.81) &  2.67 &  0.073 &  0.134 &  2.03 &  1.11 &  2.16 &  0.78 \\ 
      & 0.48  & 0.74 &  0.28 &  0.64 &  0.007 &  0.008 &  0.30 &  0.34 &  0.39 & 0.28 \\
GC30  &  2.56 &  2.75 &  2.40 &  3.94 &  0.090 &  0.197 &  2.60 &  1.78 &  (0.62) &  0.68 \\ 
 & 0.45  & 0.70 &  0.27 &  0.59 &  0.006 &  0.007 &  0.28 &  0.31 &  0.36 & 0.26 \\
GC37  &  1.49 &  0.32 &  2.03 &  2.61 &  (0.066) &  0.121 &  1.24 &  (1.97) &  1.16 &  0.95 \\ 
 & 0.50 & 0.77 &  0.29 &  0.64 &  0.006 &  0.007 &  0.30 &  0.33 &  0.37 & 0.27 \\
GC38  & 1.07 &  (1.58) &  2.95 &  (2.61) &  0.006 &  (0.091) &  1.03 &  1.19 & -0.63 &  0.49 \\  
 & 0.50 & 0.77 &  0.29 &  0.66 &  0.007  & 0.008 &  0.32 &  0.36 &  0.43 & 0.30 \\
GC45  &   1.38 & (-2.02) &  1.97 &  (5.98) &  (0.104) &  (0.144) & .... &  1.08 &  (2.91) &  0.01 \\ 
      & 0.58 &  0.88 &  0.35 &  0.75 &  0.008 &  0.009  & .... &  0.40 &  0.45 & 0.32 \\
GC47  &   (1.62) &  (0.07) &  1.26 &  5.92 &  0.114 &  0.239 &  3.46 &  2.92 &  1.91 &  1.38 \\ 
      & 0.43 &  0.65 &  0.26 &  0.55 &  0.006 &  0.007 &  0.25 &  0.28 &  0.32 & 0.23 \\
GC60  &  0.30 &  (2.68) &  (2.36) &  1.92 &  (0.040) &  0.059 &  .... &  (2.07) &  1.10 &  (1.09) \\   
      & 0.57 &  0.83 &  0.31 &  0.71 &  0.007 &  0.008 &  .... &  0.36 &  0.41 &  0.29 \\
GC74  &  -0.62 &  0.24 &  3.64 & -0.67 &  0.034 &  0.050 &  .... &  1.09 &  0.01 &  0.10 \\
      & 0.64   &  0.96 &  0.34 &  0.85 &  0.008 &  0.010 &  .... &  0.44 &  0.51 & 0.37 \\
GC78  &   2.02 &  1.19 &  2.30 &  3.19 &  0.045 &  0.144 &  1.78 &  2.75 &  1.91 &  1.27 \\ 
 & 0.63 &  0.96 &  0.36 &  0.79 &  0.008 &  0.009 &  0.38 &  0.40 &  0.45 &  0.33 \\

\hline
\end{tabular}
\label{indsred}
\end{center}
\end{table*}

We have measured the BH90 indices, and using the method outlined in
their paper, derived empirical metallicities.  These values are
obtained from an unweighted average of 6 metal-sensitive indices: CNB,
G band, Fe52, MgH, Mg$_{2}$ and $\Delta$ (a measure of the 4000 \AA\
break).  The GCs affected by internal reflections do not have CNB or
$\Delta$ included in their BH metallicity estimate.  Similarly, GCs
with poor 5202 \AA\ skyline subtraction do not have MgH or Mg$_{2}$
included.  The GCs 45 and 60 are affected by both internal reflections
and poor sky subtraction so their BH90 metallicities are based only on
the Gband and Fe52.  The final BH90 metallicity estimate is presented
in Table 4.  We also include in Table 4 the PCA metallicities derived
using the method of SB04. PCA metallicity errors include only
propagated errors in index measurement, not possible systematic errors
in the metallicity relation itself (see SB04 for details).

\subsection{Blue Horizontal Branches}

Before estimating the ages and metallicities of the GCs from
integrated spectra it is necessary to consider the effect of the
horizontal branch morphology on the indices.  Hot, luminous stars on
the blue section of the horizontal branch of a GC can make a
significant contribution to its spectrum (e.g., Lee, Yoon \& Lee
2000).  The effects of BHB stars are to both increase the intrinsic
strength of the Balmer absorption lines and raise the blue continuum
level, which in turn slightly increases the strength of all indices in
the blue.  This means that if the BHB of a GC is stronger than
modelled, younger ages and higher metallicities (due to the
age-metallicity degeneracy) are inferred.  Current SSP models do not
cover the full range of observed BHB morphology.  For example, Bruzual
\& Charlot (2003) SSPs incorporate the effect of a blue HB population
for metal-poor GCs. Thomas, Maraston \& Bender (2003; hereafter TMB03)
allow a choice of a blue horizontal branch (BHB) or red horizontal
branch (RHB) for old metal-poor populations, although this only
applies to the Balmer indices.  Thomas, Maraston \& Korn (2004;
hereafter TMK04) include BHB morphology for the old, metal-poor regime
changing to RHB morphology in the metal-rich regime. We will show that
anomalously blue HBs are likely present in several of our GCs.

\begin{figure} 
\centerline{\psfig{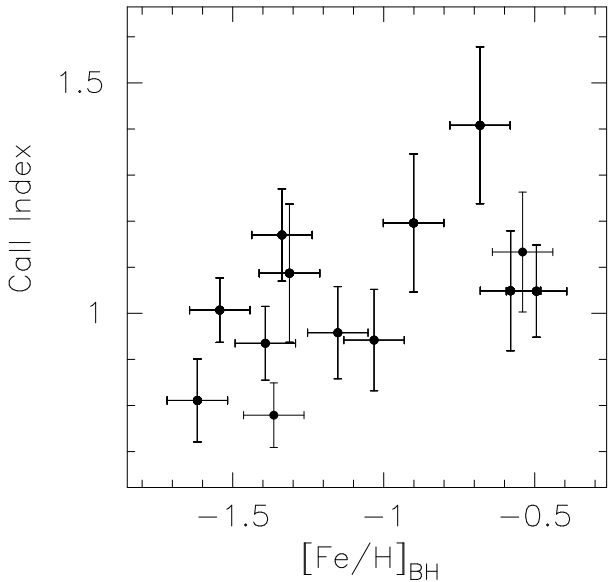}} 
\caption{The CaII index against BH90 metallicity.  There is a trend for low CaII values (indicating the presence of a blue horizontal branch) to be found in metal-poor globular clusters.  The [Fe/H]$_{BH}$ errors are indicative only.   GCs 45 and 60 are not plotted because their BH90 metallicities are based on only two indices.}
\label{caIIfe}     
\end{figure}

One indicator for the presence of BHB stars in old populations is the
Rose (1984) CaII index (Trippico 1989).  The CaII index is defined as
the ratio of the central intensity of CaII~H~+~H$\epsilon$ at
3968~\AA\ divided by the central intensity of CaII~K at 3933~\AA.  The
BHB stars greatly increase the absorption of H$\epsilon$ and therefore
lower the value of this ratio (young GCs might be expected to show a
similar behavior).  This effect, known as the ``Ca Inversion'', can be
clearly seen in GCs such as GC74 and GC38 of our sample (see Figure
\ref{spec}), which have CaII index values of 0.81 and 0.78
respectively.  Our CaII indices are presented in Table 2 (we were
unable to measure the CaII index for GC30). Although reflections may
effect the measurement of the CaII index in some GCs, the lines are
sufficiently strong so that the effect is minor. The CaII indices are
plotted in Figure \ref{caIIfe} against BH90 metallicities. This figure
shows that, as expected, GCs with BHBs tend to be metal-poor.

Problems occur with fitting the indices to SSP models when the
observed GC has a BHB contribution significantly greater than is
modelled or when BHBs are present in metal-rich GCs
(i.e. [Fe/H]$>$--1.0). Another difficulty arises for the case of
metal-poor GCs with RHBs.  Only the TMB03 SSP models cover this situation.

\subsection{Fitting Indices to Simple Stellar Population Models}

Ages and metallicities are derived by comparing the measured indices
with SSP models.  The choice of which SSP model to use is not obvious
since all the models have different advantages and limitations. We
decided to use the recent models of TMK04, who modelled the affect of
abundance ratios on the H$\gamma$ and H$\delta$ Balmer lines for the
first time. These models only differ significantly from TMB03 at high
metallicity (i.e. [Fe/H]$>$--0.35). We have also examined Vazdekis
(1999), Bruzual \& Charlot (2003) and TMB03 SSP models and find
qualitatively similar results for the derived ages, metallicities and
abundance ratios ([E/Fe]). For a comparison of SSP models see Proctor \etal (2004).

To derive accurate measures of GC parameters, particularly in view of
the problematic indices discussed in Section 3.1, we need to include
as many reliable indices as possible in the fitting process.  Here, we
use all uncontaminated indices in a multi-dimensional fitting
procedure (see PS02), using the SSP models of TMK04, which include
$\alpha$--enhancement.  A $\chi^{2}$--minimisation fit is made to the SSP
grid points, including an interpolation along all three parameter
axes.

We reject outliers by iteratively excluding the index that has the
greatest contribution to the overall $\chi^{2}$.  This decision is
based not only on the individual index's $\chi^{2}$ value, but also on
the change of the new fit obtained by excluding it.  We also require
the final fit to be stable to the further exclusion of indices (those
that do not end up stable are GCs 45, 47 and 78).  To achieve this it
was often necessary to search for specific indices that destablise the
fit.
 
Indices that are excluded on this basis are in parentheses in Tables 2
and 3. On average, after all the exclusions, approximately 2/3 of the
indices are used in the final fit.  A major advantage of this method
is that we are not reliant on Balmer indices (which can be
contaminated by galaxy emission or strongly affected by BHBs).  The
errors given for the final derived parameters are statistical $1 \sigma$ confidence 
intervals calculated by a Monte Carlo style method
with 1000 realisations of the best-fit SSP.

\section{Derived GC Ages, Metallicities and Alpha Abundances}\label{sec_aam}

The results of our fitting procedure are given in Table 4.  We also
present the metallicity estimates by the methods of BH90 and SB04.
For GCs 45, 47 and 78, the SSP-derived parameters should be regarded
with some caution as the final values are less reliable than those for
the rest of the sample.  

\begin{table*}
\begin{center}
\renewcommand{\arraystretch}{1.5}
\caption{\scriptsize Globular Cluster derived parameters.  Age, [Fe/H]$_{SSP}$, [E/Fe], [Z/H], [Fe/H]$_{BH}$ and [Fe/H]$_{PCA}$ for the 16 GCs.  Errors given are 1 $\sigma$ statistical errors, the PCA errors only include index errors and thus may be underestimates.  Values in brackets indicate unreliable BH90 metallicity estimates.}
\begin{tabular}{lrccccc}
\hline
\hline
ID  & Age  & [Fe/H]$_{SSP}$ & [E/Fe] & [Z/H] & [Fe/H]$_{BH}$ & [Fe/H]$_{PCA}$\\
GC & (Gyr) & (dex) & (dex) & (dex) & (dex) & (dex)\\
\hline
GC7  & 15.0$\pm$5.4  &   -1.27$\pm$0.18   &  0.34$\pm$0.16   &  -0.95$\pm$0.15    & -1.54 & -1.32$\pm$0.08\\
GC9  & 13.3$\pm$3.5  &   -0.74$\pm$0.13   &  0.15$\pm$0.10   &  -0.60$\pm$0.11    & -0.68 & -0.38$\pm$0.07\\
GC12 & 12.6$\pm$4.2  &   -0.63$\pm$0.12   &  0.38$\pm$0.09   &  -0.28$\pm$0.08    & -0.49 & -0.21$\pm$0.07\\
GC17 & 11.9$\pm$2.0  &   -1.58$\pm$0.23   &  0.32$\pm$0.22   &  -1.28$\pm$0.07    & -1.39 & -1.87$\pm$0.08\\
GC18 & 15.0$\pm$3.8  &   -0.65$\pm$0.11   &  0.34$\pm$0.08   &  -0.33$\pm$0.08    & -0.58 & -0.33$\pm$0.08\\
GC19 & 10.6$\pm$1.8  &   -1.38$\pm$0.22   &  0.06$\pm$0.22   &  -1.33$\pm$0.08    & -1.34 & -1.16$\pm$0.09\\
GC28 & 15.0$\pm$6.6  &   -0.81$\pm$0.21   &  0.12$\pm$0.13   &  -0.70$\pm$0.18    & -0.90 & -0.81$\pm$0.09\\
GC29 & 15.0$\pm$7.3  &   -1.15$\pm$0.23   &  0.27$\pm$0.13   &  -0.90$\pm$0.17    & -1.15 & -1.22$\pm$0.09\\
GC30 & 15.0$\pm$3.1  &   -0.63$\pm$0.11   &  0.30$\pm$0.07   &  -0.35$\pm$0.07    & -0.29 & -0.53$\pm$0.09\\
GC37 & 12.6$\pm$2.9  &   -1.42$\pm$0.19   &  0.34$\pm$0.17   &  -1.10$\pm$0.09    & -1.03 & -1.10$\pm$0.09\\
GC38 & 8.9$\pm$2.0   &   -2.05$\pm$0.17   &  0.50$\pm$0.07   &  -1.58$\pm$0.16    & -1.36 & -1.91$\pm$0.10\\
GC45 & 15.0$\pm$7.1  &   -2.38$\pm$0.46   &  0.80$\pm$0.37   &  -1.63$\pm$0.40    &(-1.44)& -0.95$\pm$0.11\\
GC47 & 15.0$\pm$3.8  &   -0.24$\pm$0.12   &  0.15$\pm$0.06   &  -0.10$\pm$0.09    & -0.54 & -0.37$\pm$0.08\\
GC60 & 10.6$\pm$2.1  &   -2.27$\pm$0.24   &  0.50$\pm$0.19   &  -1.80$\pm$0.16    &(-0.79)& -1.79$\pm$0.10\\
GC74 & 13.3$\pm$4.3  &   -2.08$\pm$0.42   &  0.30$\pm$0.41   &  -1.80$\pm$0.18    & -1.62 & -1.84$\pm$0.11\\
GC78 & 10.6$\pm$4.1  &   -0.70$\pm$0.21   &  0.03$\pm$0.14   &  -0.68$\pm$0.14    & -1.31 & -0.85$\pm$0.11\\
\hline
\end{tabular}
\label{tablefits}
\end{center}
\end{table*}

\subsection{Age}

The derived age for our sample are given in Table 4. All GCs have ages
of greater than 10 Gyr; the only exception is GC38, with an inferred
age of 8.9 Gyr.  Our previous discussion of the effects of BHBs on
age estimation in Section 3.2 leads us to suggest that GC38, with the
strongest BHB in our sample, is actually older than its inferred age.  Assuming that
the BHB in GC38 is stronger than modelled, then the true age and
metallicity of this GC will be older and more metal-poor than given
(although the BHB almost exclusively affects age-sensitive
indices, age and metallicity are degenerate parameters in the SSP
fitting process).

\begin{figure} 
\centerline{\psfig{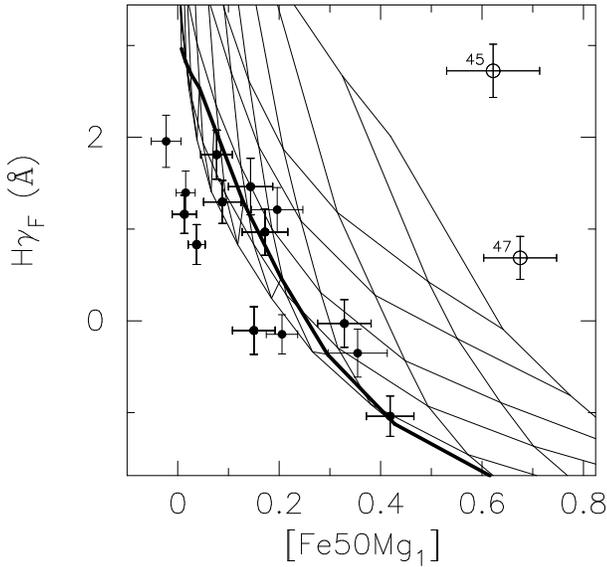}} 
\caption{Index-index plot of H$\gamma_F$ vs [Fe50Mg$_1$].  Open
circles indicate GCs for which one or more of the plotted indices are
significantly affected by emission features in the index side-bands
(i.e. GCs 45 and 47).  Filled circles indicate GCs for which the three
indices give fits similar to the full multi-index solution.  The TMK04
grid lines shown are for [E/Fe]=+0.3 with metallicity in 0.25 dex
steps from -2.25 to +0.5 (left to right) and ages of 1,2,3,5,8,12 and
15 Gyr (top to bottom), the heavier line is 15 Gyr. The GCs with
reliable indices are consistent with old ages and a range of
metallicities. Note that the ages used in this work are not derived from this plot.}
\label{indexplot}     
\end{figure}

An index-index plot is presented in Figure \ref{indexplot} of
H$\gamma_F$ vs [Fe50Mg$_1$], where [Fe50Mg$_1$] = Fe5015 x
Mg$_1$. We have chosen this combination of Balmer and
metal-sensitive lines as these indices appear to be the most accurate in our GC sample.
The plot indicates that the GCs are consistent with being very old
and covering a range of metallicity typical for GC systems. The GCs for which one of these
two indices are not reliable, and hence not used in the best-fit solution,
are also shown. These are GCs 45 and 47, for 
which Fe5015 and H$\gamma_F$ are affected by emission features in the index side-bands.
These GCs have the most galactic background light due to their spatial positions (see Figure \ref{positions}).

\begin{figure} 
\centerline{\psfig{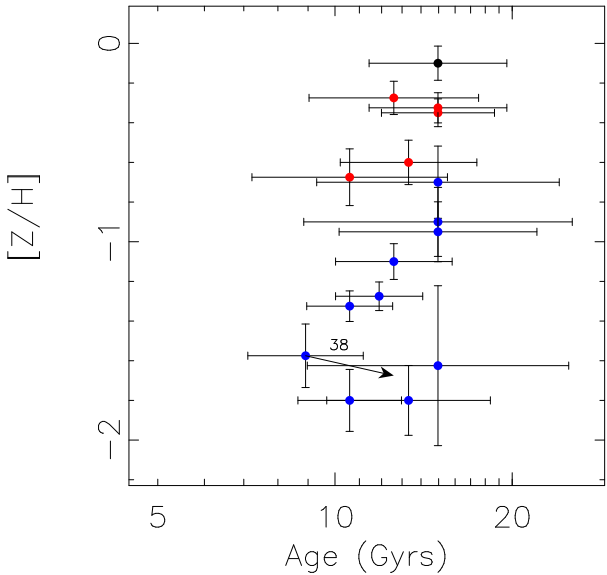}} 
\caption{Age-metallicity distribution. The GCs are consistent with a coeval age of $\sim 13$ Gyr, and show a range of metallicities of 
$\sim 2$ dex.  The arrow on GC38 indicates the direction it would shift towards assuming it is older than the SSP age indicates (due to the 
presence of a BHB artificially making it appear younger). In colour versions of this plot, the GCs are colour-coded according to their 
observed colour. The most metal-rich GC has an anomalously blue colour, otherwise the metal-poor GCs have blue colours and the metal-rich GCs are red.}
\label{agez}     
\end{figure}

Figure \ref{agez} shows an age-metallicity plot for the GCs with
values from Table 4. The GCs all have old ages for a large range in
total metallicity ([Z/H]).  The BHB GC38 has an arrow on it pointing in
the direction corresponding to the age-metallicity 3:2 degeneracy
(Worthey 1994).

\subsection{Metallicity}

\begin{figure} 
\centerline{\psfig{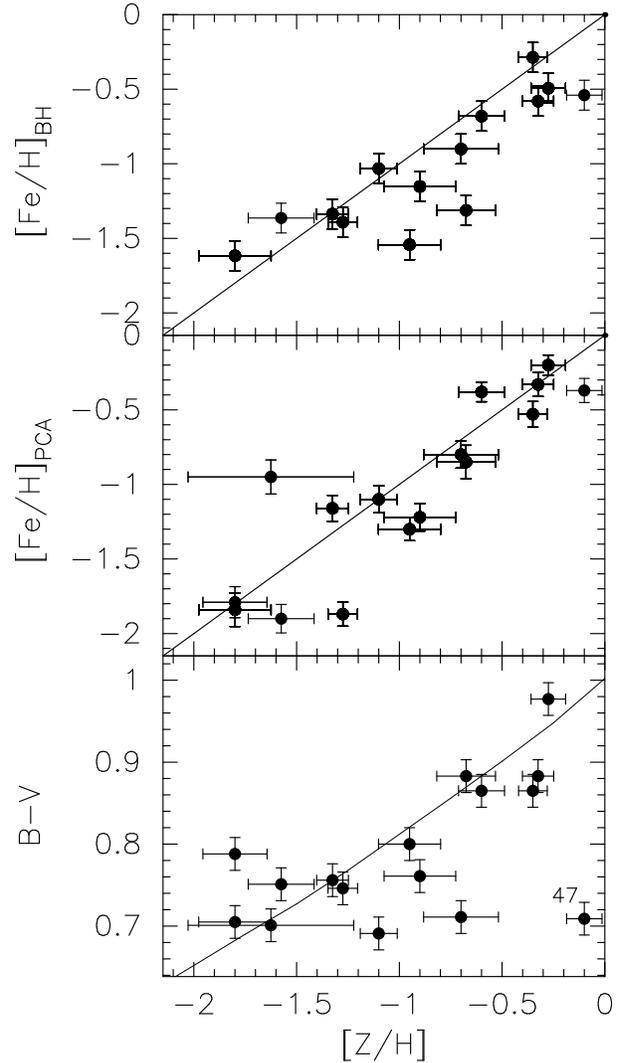}}
\caption{Comparison of empirical metallicities and observed colours
with total metallicity.  The top two plots show the metallicity from
the method of BH90 ([Fe/H]$_{BH}$) and from SB04 ([Fe/H]$_{PCA}$). A
one-to-one line is also shown.  
GCs 45 and 60 which have unreliable BH90 metallicities are not
plotted. A good correlation is seen with points lying close to the
one-to-one line for both empirical methods. The lower plot shows the
observed B--V colour from Forbes \etal (2001b) with [Z/H]. A reasonable
agreement is found with the 12 Gyr isochrone from the TMK04 SSP models
that include BHBs (solid line).  The major outlier GC47 (which has an
inexplicably blue colour) is labelled.}
\label{metal}     
\end{figure}

A test of the reliability of the derived total metallicity [Z/H] is
obtained by plotting this measure against the empirical metallicity
derived by the method of BH90 and that derived by the PCA method of
SB04 (see Table 4).  For the old ages of these
GCs both methods are well-calibrated. Figure \ref{metal} shows there
is generally good agreement between the different methods over a 
metallicity range of $\sim$2 dex. The PCA metallicities appear to show
somewhat better agreement than those of BH90 with [Z/H] derived from SSP models.

As a further consistency check of our results we plot the observed
B--V colour from Forbes \etal (2001b) versus total metallicity [Z/H] in
Figure \ref{metal}. We can see that the metallicity correlates well with
the observed colour and is consistent with the 12 Gyr isochrone of the TMK04 SSP 
models with BHBs.
We believe GC47 to have an anomalous colour (see discussion in the following paragraph). 

\begin{figure} 
\centerline{\psfig{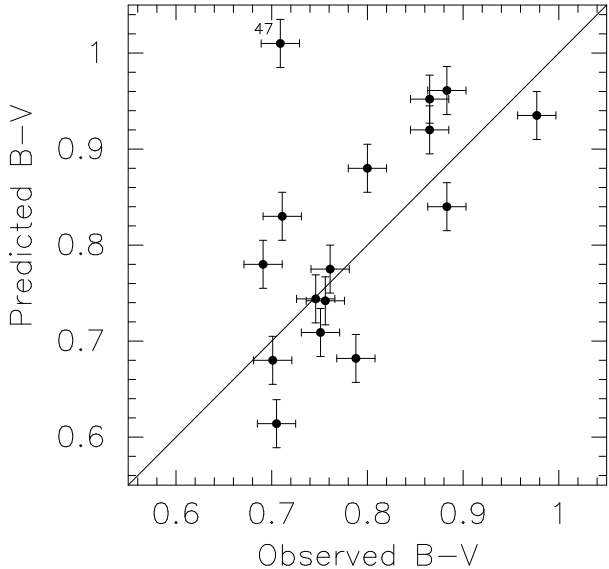}} 
\caption{The B--V colour predicted from the inferred [Z/H] and age 
is plotted against the observed B--V colour from Forbes \etal (2001b).  
The errors for the predicted colours are indicative only.  
GC47 which has an anomalous colour is labelled. For the other GCs there is good agreement between the predicted and observed colours.}
\label{colourpred}     
\end{figure}

Finally, we compare the observed colours of these GCs with colours
predicted from the SSP derived ages and total metallicities.  For GCs
with [Z/H]$\le$--1 we use the BHB mass loss models and for [Z/H]$>$--1
we use the RHB models with no mass loss.  This is consistent with the
HB morphology suggested by the CaII index values.  The predicted
colour errors are indicative estimates only.  Figure \ref{colourpred}
shows that there is reasonable agreement between the predicted and
observed colour. The main outlier is GC47. The colours measured for
GC47 disagree with its inferred age and metallicity.  Also the spectra
(see Figure \ref{spec}) suggests that GC47 should have a colour
similar to the reddest GC (i.e. GC12).  Thus we suspect a systematic
error in the observed colour for GC47.

\subsection{Abundance Ratio}

\begin{figure} 
\centerline{\psfig{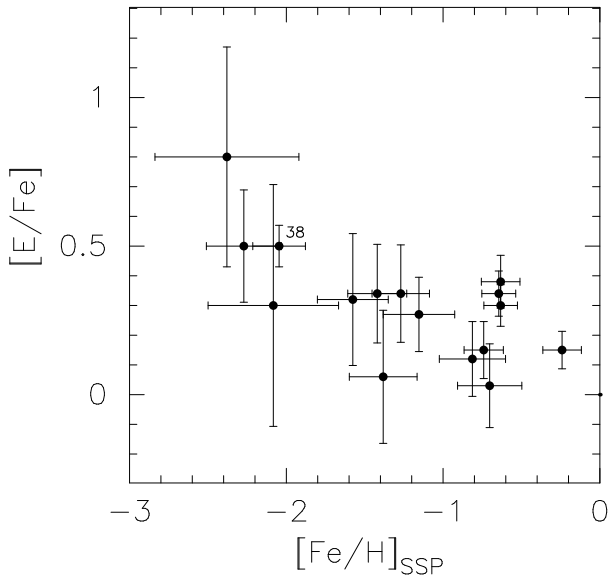}} 
\caption{Abundance ratio against SSP derived iron metallicity.  The GCs show a trend of decreasing [E/Fe] with increasing [Fe/H]$_{SSP}$.  However within errors a constant value of twice solar ([E/Fe]=+0.3) is consistent with the data.}
\label{alphafeplot}     
\end{figure}

We show relationship between [E/Fe] and [Fe/H]$_{SSP}$ in Figure
\ref{alphafeplot}. The sum of these two measures approximates the
total metallicity [Z/H] ([Fe/H]$_{SSP}$ is derived from the fitted
value of [Z/H]). The plot shows that [E/Fe] is consistent with
a value of twice solar ([E/Fe]=+0.3), except perhaps at the lowest
metallicities ([Fe/H]$_{SSP}<$--2) where the extrapolation of the SSP model
to high [E/Fe] is becoming unreliable. Such values are similar to
those found for Galactic GCs (e.g Proctor \etal 2004).

\section{Galaxy Spectrum}\label{sec_long}

\subsection{Long-slit Observations}

On the same observing run (see Section \ref{sec_obs}) a 300s long-slit
exposure along a P.A. = 60$^{o}$ centred on NGC~1052 itself was
obtained.  Standard reduction methods were used with IRAF software.
Two dimensional distortions were measured on the arc lamp frames then
removed using the task transform. The width for each aperture
extracted increases with radius to achieve a signal-to-noise of $\sim
30$ at 5000 \AA\ for most of the apertures.  The radial extent over
which we can measure useful spectra is limited to $\sim~18$''
west-south-west of the galactic centre and $\sim~7.5$'' in the
east-north-east direction.

For each spectrum, recession velocities were measured by
cross-correlation using fxcor in IRAF.  Velocity dispersions for the
galaxy apertures were measured by cross-correlation with standard
stars and comparison of the FWHM velocities with cross-correlations of
the standard stars broadened by a range of velocity dispersions.  This
kinematic data is presented in Table 5.

\subsection{Derived Parameters}

\begin{figure} 
\centerline{\psfig{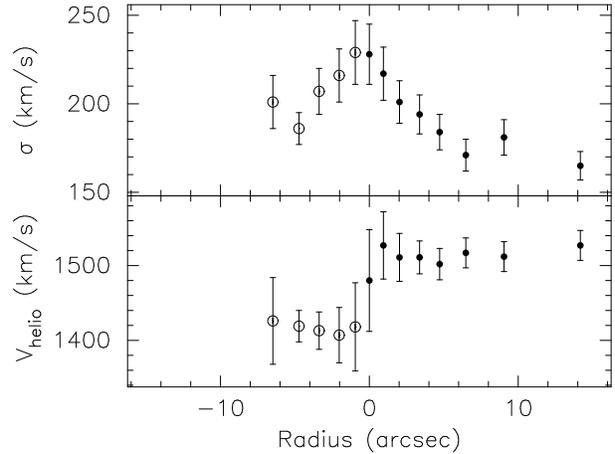}}
\caption{Recession velocity and velocity dispersion as functions of radius.  Negative radii shown as open symbols indicate east-north-east of the galaxy centre, filled symbols indicate west-south-west.  Rotation of order 50 km/s about the systemic velocity of 1470 km/s is clear from the lower panel.  The velocity dispersion shown in the upper panel is peaked towards the galaxy centre.}
\label{rotate}     
\end{figure}

Figure \ref{rotate} shows that the galaxy rotates on the order of 50
km/s along our observed axis of P.A.= 60$^{o}$. This rotation is
consistent with that measured by Binney, Davies \& Illingworth (1990),
who find rotation of $\sim$100 km/s along the major axis (P.A.=
120$^{o}$) and none along the minor axis.  Our observed axis is at an
angle of 60$^{o}$ to the major-axis, therefore we expect the magnitude
of rotation to be about cos(60$^{o}$)=0.5 of that found along the
major-axis as observed.  The velocity dispersion is centrally peaked,
with a maximum value of $\sim$230 km/s, as found by Binney \etal
(1990).

\begin{table*}
\begin{center}
\caption{\scriptsize Galaxy parameters. Radius is the centre of each aperture, where negative radii indicate east-north-east of the galaxy centre. 1 $\sigma$ errors are presented.}
\renewcommand{\arraystretch}{1.5}
\begin{tabular}{ccccrrr}
\hline
\hline
Radius & V$_{helio}$ & $\sigma$ & Age & [Fe/H] & [E/Fe] & [Z/H]\\
(arcsec)  & (km/s) & (km/s) & (Gyr) & (dex) & (dex) & (dex)\\
\hline
14.1 & 1527$\pm$20 & 165$\pm$ 8   & 4.7$\pm$3.0 & -0.08$\pm$0.22  & 0.42$\pm$0.07 & 0.32$\pm$0.21 \\
9.1  & 1512$\pm$20 & 181$\pm$10   & 4.5$\pm$2.9 &  0.15$\pm$0.24  & 0.21$\pm$0.09 & 0.35$\pm$0.28 \\
6.5  & 1517$\pm$20 & 171$\pm$ 9   & 1.9$\pm$1.2 &  0.48$\pm$0.24  & 0.36$\pm$0.07 & 0.81$\pm$0.23 \\
4.7  & 1502$\pm$21 & 184$\pm$10   & 1.8$\pm$0.9 &  0.65$\pm$0.17  & 0.42$\pm$0.06 & 1.05$\pm$0.19 \\
3.4  & 1511$\pm$22 & 194$\pm$11   & 1.5$\pm$0.6 &  0.63$\pm$0.18  & 0.57$\pm$0.06 & 1.16$\pm$0.18 \\
2.0  & 1511$\pm$32 & 201$\pm$12   & 2.1$\pm$1.0 &  0.65$\pm$0.13  & 0.51$\pm$0.05 & 1.13$\pm$0.15 \\
1.0  & 1527$\pm$45 & 217$\pm$15   & 1.6$\pm$0.4 &  0.55$\pm$0.14  & 0.81$\pm$0.04 & 1.31$\pm$0.15 \\
0.0  & 1480$\pm$68 & 228$\pm$17   & ....        & ....       & ....        &    ....     \\			  
-1.0 & 1418$\pm$59 & 229$\pm$18   & 1.8$\pm$0.4 & 0.65$\pm$0.11  & 0.84$\pm$0.05 & 1.44$\pm$0.12 \\
-2.0 & 1407$\pm$37 & 216$\pm$15   & 1.6$\pm$0.4 & 0.65$\pm$0.11  & 0.69$\pm$0.04 & 1.30$\pm$0.12 \\
-3.4 & 1413$\pm$25 & 207$\pm$13   & 1.9$\pm$0.7 & 0.63$\pm$0.13  & 0.51$\pm$0.05 & 1.11$\pm$0.15 \\
-4.7 & 1419$\pm$21 & 186$\pm$ 9   & 2.4$\pm$1.3 & 0.53$\pm$0.16  & 0.36$\pm$0.07 & 0.86$\pm$0.18 \\
-6.5 & 1426$\pm$58 & 201$\pm$15   & 1.9$\pm$0.7 & 0.65$\pm$0.13  & 0.36$\pm$0.06 & 0.99$\pm$0.15 \\   
\hline
\end{tabular}
\label{tablefits}
\end{center}
\end{table*}

\begin{figure}
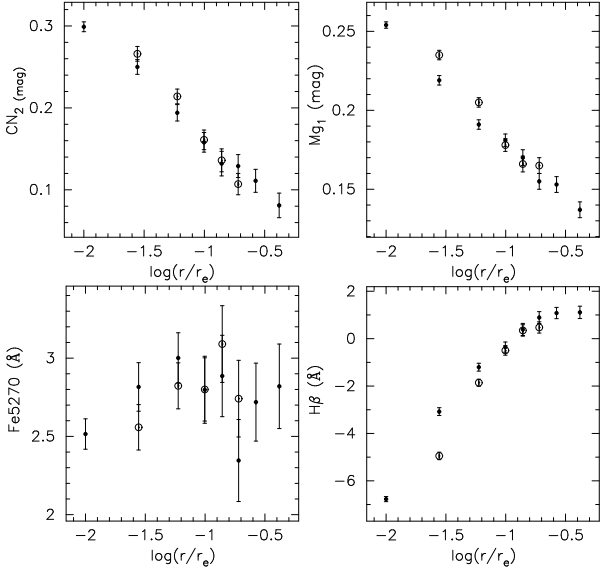
 
\centerline{\psfig{figure=fig11a.ps,width=0.22\textwidth,angle=270}  \psfig{figure=fig11b.ps,width=0.22\textwidth,angle=270}}
\centerline{\psfig{figure=fig11c.ps,width=0.22\textwidth,angle=270}  \psfig{figure=fig11d.ps,width=0.22\textwidth,angle=270}}  
\caption{Representative plots of indices vs. log radius in terms of the effective radius (r$_e$=34''). We have folded the data about the centre with points from the east-north-east marked with open symbols, filled symbols indicate west-south-west.  The strong emission in H$\beta$ is clearly seen as negative values.  This highlights the need to exclude the Balmer indices from our fitting process.  The Mg$_{1}$ gradient when viewed in tandem with the Fe5270 lack of gradient implies that [Fe/H] is roughly constant with radius and that the abundance ratio is much greater in the central region.  The gradient in other $\alpha$--element sensitive indices reinforces this picture.}
\label{radial}     
\end{figure}

Lick indices were measured for the galaxy at each radius and the same
offset to the Lick system was applied as per the GC data.  Denicolo
\etal (2005) presents indices, measured along the major axis of
NGC~1052, which are consistent with our observed values. Several
indices show gradients with galactocentric radius.  A sample of these
are plotted in Figure \ref{radial}, showing some clear radial trends.
The metallicity and [E/Fe] sensitive indices (e.g., CN$_{2}$ and
Mg$_{1}$) have a strong gradient towards higher values in the centre.
A similar trend is observed for other [E/Fe] sensitive indices, e.g.,
Mgb.  The Fe5270 line, which is primarily sensitive to iron
metallicity [Fe/H], shows almost no gradient.  Other Fe lines such as
Fe4383, Fe4531 and Fe5406 also show no gradient with radius.  These
trends are also evident in the Denicolo \etal (2005) data.  The galaxy
therefore posses strong gradients in [E/Fe] sensitive indices, but no
gradients in the [Fe/H] sensitive indices.  The effect of central
galactic emission is obvious in the H$\beta$ index plot, with large
negative values at small radii.  The other Balmer indices and Fe5015
are similarly affected by emission.

\begin{figure} 
\centerline{\psfig{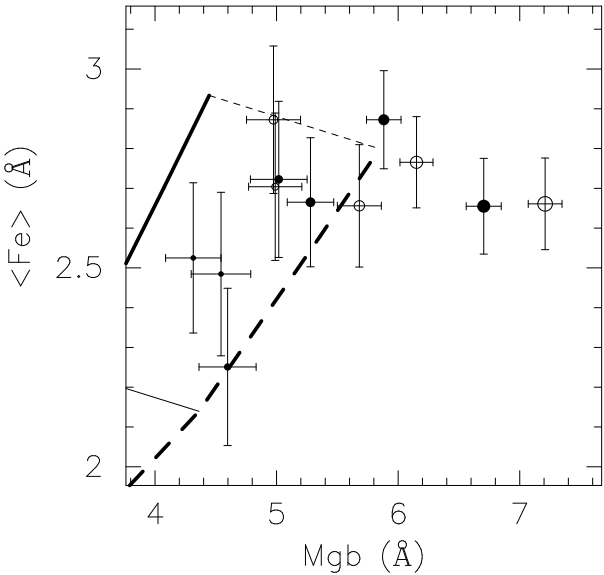}} 
\caption{Plot of $<$Fe$>$ vs Mgb indices. Bold lines shown the 
2 Gyr isochrones of [E/Fe]=+0.3 dex (solid) and [E/Fe]=+0.6 dex (dashed). 
The thin lines show [Fe/H]=+0.0 (solid) and +0.4 (dashed).  
Point sizes decreases with radius. The open symbols are the east-north-east 
apertures and the filled symbols indicate the apertures towards 
the west-south-west. 
The plot shows that the central apertures are the most $\alpha$--element enhanced 
with abundance ratios above +0.6 dex.}
\label{mgbfe}     
\end{figure}

Due to the high total metallicity of the galaxy, we use the Fe--
method from PS02 as opposed to the Trager \etal (2000a) method used
earlier to calculate the [E/Fe] .  This is necessary because the
isochrone shape is driven by [Z/H] at low metallicity and [Fe/H] at
high metallicity (see PS02 for a full explanation).  Due to the high
$\alpha$--element abundance ratios present in the galaxy, we
extrapolate the TMK04 models to [E/Fe]=+0.9.  Figure \ref{mgbfe}
demonstrates the extremely high abundance ratios present in the
central regions by showing a plot of $<$Fe$>$ vs Mgb, where
$<$Fe$>$=(Fe5270+Fe5335)/2.

To obtain reasonable fits to the TMK04 SSP models it was necessary to
exclude several indices from all apertures during the fitting process.
Galactic emission which increases towards the centre, forces the
removal of all the Balmer lines and Fe5015 from the fitting process.
The index C4668 was also removed due to its systematic deviance from
the best fit.

\begin{figure}
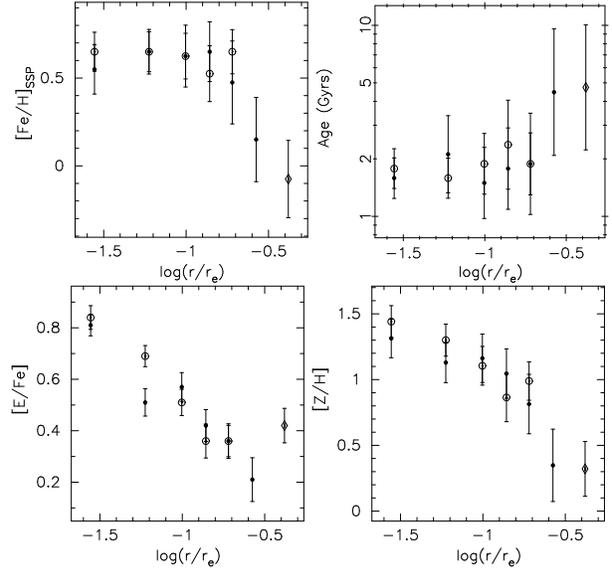
 
\centerline{\psfig{figure=fig13a.ps,width=0.22\textwidth,angle=270}    \psfig{figure=fig13b.ps,width=0.22\textwidth,angle=270}}  
\centerline{\psfig{figure=fig13c.ps,width=0.22\textwidth,angle=270}  \psfig{figure=fig13d.ps,width=0.22\textwidth,angle=270} } 
\caption{The derived stellar population parameters for the galaxy field stars are plotted with radius.  We have folded the data about the galaxy centre with points from the east-north-east marked with open circles, filled symbols indicate west-south-west.  The open diamond indicate the aperture which has a less reliable fit.  The central galaxy aperture is not shown.  There is no strong gradient in either age or iron metallicity. There is a clear trend of $\alpha$--element abundance ratio increasing towards the centre.  Total metallicity shows a radial gradient.}
\label{longall}     
\end{figure}

We also clipped individual deviant indices following a similar method
to Section 3.3.  This resulted in a total of 12\% of the remaining
indices being removed (mostly Mg$_{2}$, Ca4455 and Fe4383).  We were
unable to obtain reasonable fits for the central aperture, probably
due to strong galactic emission, and do not present the fitted
parameters for it.  The values obtained for age, [Fe/H], [E/Fe] and
[Z/H] are presented in Table 5 and also plotted against radius in
Figure \ref{longall}.

Figure \ref{longall} shows no strong radial gradient in either age or
[Fe/H]. There is however a very strong gradient of
decreasing [E/Fe] with increasing radius.  In the central
arcsecond the estimated [E/Fe] of $\sim$+0.8 dex is unusually
high.

\section{Globular Cluster Kinematics}\label{sec_gal}

Using the kinematic information from the 16 GCs, we can estimate the
mass enclosed within the radius of the GC system observed.  We use the
projected mass estimator (Evans \etal 2003), assuming isotropy and an
r$^{-4}$ distribution to derive a mass of 1.7$\pm 0.9 \times 10^{12}
M_{\sun}$ within 19 kpc ($\sim 6.5$ r$_{e}$). The mass estimate error
was calculated by bootstrapping the observed velocities and
errors. van Gorkom \etal (1986) used HI kinematics to measure a mass
of $3.1 \times 10^{11} M_{\sun}$ within 23 kpc.  We have limited spatial
coverage and too few GCs to make any comment on systematic rotation of
the GC system. Since our mass estimate is significantly higher than
that for the HI kinematics, our assumption of isotropy may be
incorrect. For example, if the orbits
had a rather strong tangental bias of $\sigma_t$/$\sigma_r$ = 2, then
our tracer mass estimate would be 30\% too large (Evans et al. 2003).

\section{Discussion}\label{sec_disc}

To summarise, we have obtained spectra for 16 GCs which sample both
the blue and red sub-populations in NGC~1052.  Ages, metallicities and
abundance ratios for these GCs have been derived using the multi-line
fitting technique of PS02 applied to the SSPs of TMK04.  We find that
the blue GCs in NGC~1052 to be uniformly old $\geq$10 Gyr and
metal-poor (with the exception of GC47 for which we strongly suspect
an incorrect colour). The red GCs are similarly old but metal-rich.
We do not find any young GCs associated with the recent merger event,
even though we have sampled the expected colour and luminosity regime
for young GCs.

We do, however find evidence for young \emph{galaxy} stars associated
with the merger event. Our luminosity-weighted age of $\sim 2$ Gyr for
the NGC 1052 stars is compatible with that estimated from the HI tidal
tails and infalling HI gas of $\sim 1$ Gyr (van Gorkom \etal 1986).

The high central value and strong gradient observed for [E/Fe] in NGC
1052 is unusual for an elliptical galaxy. For example, Mehlert \etal
(2003) find negligible gradients in [E/Fe] for 35 early-type galaxies
in their sample of Coma galaxies. Proctor (2002) using the same method
as this work finds a maximum [E/Fe]$\simeq$+0.4 and no significant
[E/Fe] gradient in the bulges of 28 late and early-type galaxies.

The $\alpha$--element abundance ratio for NGC~1052 is largely
insensitive to the combination of indices we fit, unlike age and
metallicity.  Therefore, barring a major contamination of all the
[E/Fe] sensitive indices (which must also affect the observations of
Denicolo \etal 2005) we can confidently say that the central galaxy
starlight is more $\alpha$--element enhanced than $\sim +0.6$
dex. Such $\alpha$--element enhancements and the deduced high central
metallicities ([Z/H]=+1.3 dex) are far beyond current SSP models.
Therefore, it is impossible to pin down exact values for the central
galaxy field stars.  At [Z/H]=+1.3 the total metallicity would be 20
times that of the Sun, implying a metal content by mass of $\sim$40\%.
This value is unrealistically high, which suggesting that our
extrapolations beyond the current SSP models are unreliable.

van Gorkom \etal (1986) speculated that NGC~1052 had accreted a
gas-rich dwarf galaxy about 1 Gyr ago. They find a total HI mass of
$\sim 5 \times 10^8 M_{\sun}$ which seems consistent with this
idea. Such a small quantity of gas could also explain why we have
detected no GCs forming in this merger (even allowing for some cold
gas to have been used up in field star formation and/or ionised). Even
a small burst by mass will give a young age for the galaxy central
stars (Terlevich \& Forbes 2002).

Our results also indicate that the central merger-induced starburst
involved very high metallicity and $\alpha$--element enhanced
gas. Such high metallicity gas is not usually associated with dwarf
galaxies but massive spirals. The most obvious explanation for the
high $\alpha$--element enhancement is a high central concentration
Type II supernovae from the starburst. Inflows over any significant
timescale are less likely, since no radial metallicity gradient is
seen. A spiral galaxy with a small bulge and on-going star formation
in the disk could produce the young central ages observed without star
formation from the merger itself being the dominant luminosity source.

To better evaluate these possibilities we need stellar population
information for the galaxy out to at least one effective radius.  This
will provide an age, metallicity and $\alpha$--element abundance ratio for the
assumed old ($>$10 Gyr) underlying population.  With this information the next
step would be to apply a chemo-gas code in an attempt to reproduce the
observed properties.

In terms of the GC formation models described in the Introduction, the
detection of old red and blue GC sub-populations in NGC~1052 is
generally consistent with the predictions of both Forbes \etal (1997)
and Cote \etal (1998). The ages are not yet accurate enough to
determine whether the red GCs are a few Gyr younger than the blue ones
as expected in Forbes \etal (1997).  On the other hand, we find no
evidence for GCs forming in the merger event, in contrast to the
expectations of the merger model of Ashman \& Zepf (1992).

There are now several examples of galaxies with spectroscopically
confirmed {\it minor} contributions to their GC systems from recently
formed GCs, while the vast bulk of their GCs appear to be very old.
NGC 1399, a central cluster galaxy which is old and has no photometric
indication of intermediate-age GCs, may have a small number of $\sim$2
Gyr old GCs (Forbes \etal 2001a). Photometry of GCs in the merger
remnant NGC 3610, which has a spectroscopic age of 1.6$\pm$0.5 Gyrs
(Denicolo \etal 2005), suggested the presence of an intermediate-age
sub-population (Whitmore \etal 2002). However, spectroscopy of some of
these candidate intermediate-age GCs indicates that only a small
portion of these clusters are in fact $\sim$2 Gyr old (Strader \etal
2004). 

\begin{figure} 
\centerline{\psfig{figure=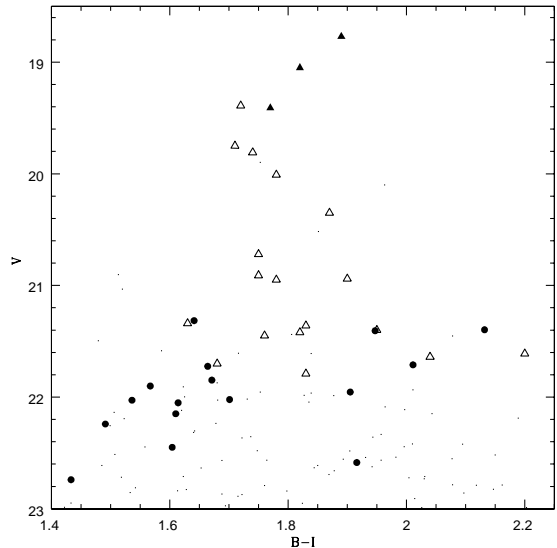,width=0.45\textwidth,angle=0}}  
\caption{The colour magnitude diiagram for spectroscopically confirmed GCs around NGC~1052 (filled circles).  The small dots are GC candidates around NGC~1052 from the imaging of Forbes \etal (2001b).  The triangles are spectroscopically confirmed GCs around NGC~1316 (Goudfrooij \etal 2001) with their magnitude's adjusted to account for the distance modulus difference.  Only the filled triangles had high enough S/N for ages and metallicities to be measured, the open triangles have measured velocities but not indices.  Most of the NGC~1316 GCs have intermediate colours (B--I$\sim$1.8) and are significantly brighter than the NGC~1052 GCs.  Therefore we do not expect a similar sub-population of $\sim$3 Gyr GCs in NGC~1052.}
\label{disc1316}     
\end{figure}

Perhaps the best example of intermediate-age GCs in a merger remnant
is NGC~1316, it has a similar age ($\sim$3 Gyrs) and lies at a similar
distance (22.9 Mpc) to NGC~1052 (Goudfrooij \etal 2001).  Figure
\ref{disc1316} shows a comparison of the colour magnitude diagrams for
the spectroscopically confirmed GCs of these two galaxies.  Only the 3
brightest NGC~1316 GCs had high enough S/N to measure ages, and these
are 2 magnitudes brighter in V than the brightest GCs for which we
obtained spectra in NGC~1052.  All three GCs were found to be $\sim$3
Gyrs old and solar metallicity (Goudfrooij \etal 2001).  The NGC~1316
GC colour distribution peaks at B--I$\sim$1.8, which lies between the
peaks of the NGC~1052 GC colour distribution.  There is no significant
population of NGC~1052 GCs in the same colour and magnitude parameter
space as the young NGC~1316 GCs. Therefore we do not expect a similar
sub-population of $\leq$3 Gyr GCs in NGC~1052.

\section{Conclusions}\label{sec_conc}

We have obtained low-resolution spectra for 19 Globular Cluster (GC)
candidates associated with the merger remnant elliptical NGC~1052.  Of
this sample, 16 are identified as $\it{bona~fide}$ GCs by their radial
velocities. Using these velocities, we derive a virial mass of
1.7$\pm$0.9 x 10$^{12} M_{\sun}$ within a radius of $\sim$6.5 r$_{e}$
(19 kpc).  If the orbits had a strong tangental bias then
our tracer mass estimate would be 30\% too large.

Using the multi-index $\chi^{2}$ fitting technique of Proctor \&
Sansom (2002), and the simple stellar population models of Thomas,
Maraston \& Korn (2004) we derive individual ages, metallicities and
abundance ratios for the 16 confirmed GCs.  We find all of the GCs to
be very old, i.e. $\geq 10$ Gyr, with a range
of metallicities. The predicted colours, based on the derived ages and
metallicities, agree well with the observed GC colours. We find no
evidence for young GCs associated with the likely minor merger event
$\sim$1 Gyr ago.  

We also obtained a long-slit spectrum covering the central $\sim 15$''
of NGC~1052. No strong gradient in either age or metallicity was
found.  However, a large abundance ratio gradient exists.  The stellar
population in the central regions of NGC~1052 has a
luminosity-weighted age of $\sim 2$ Gyr with [Fe/H]$\sim +0.6$ and a
very high $\alpha$--element abundance ratio of $\sim$+0.8 dex. The
recent central star formation episode was most likely induced by
infalling gas associated with the recent merger. Thus, although
NGC~1052 shows substantial evidence for a recent merger and an
associated starburst, it appears that the merger did not induce the
formation of many, if any, new GCs. As the formation of luminous star
clusters appears to accompany most significant star formation events,
the absence of young GCs and the high [E/Fe] values in the center of
the galaxy (suggesting short star formation timescales) may indicate
that a relatively small amount of star formation occurred in the
merger. This interpretation is consistent with ``frosting'' models for
the formation of early-type galaxies (e.g., Trager \etal 2000b).

\section{Acknowledgments}\label{sec_ack}

We thank Soeren Larsen for help preparing the slit mask.  Part of this
research was funded by NSF grant AST-02-06139 The data presented
herein were obtained at the W.M. Keck Observatory, which is operated
as a scientific partnership among the California Institute of
Technology, the University of California and the National Aeronautics
and Space Administration.  The Observatory was made possible by the
generous financial support of the W.M. Keck Foundation. This research
has made use of the NASA/IPAC Extragalactic Database (NED), which is
operated by the Jet Propulsion Laboratory, Caltech, under contract
with the National Aeronautics and Space Administration. DF and RP
thanks the ARC for their financial support.

\end{document}